\documentclass[%
 reprint,
showpacs,
 amsmath,amssymb,
 aps,
]{revtex4-1}

\usepackage{graphicx}
\usepackage{dcolumn}
\usepackage{bm}

\begin{document}

\title{Electromagnetic Response  for Modulation at Optical Time Scale.}
\author{Evgenii E. Narimanov}
\affiliation{School of Electrical and Computer Engineering and Birck Nanotechnology Center, Purdue University, West Lafayette, Indiana 47907, USA}
\date{\today}

\begin{abstract}
Rapid modulation of the electromagnetic response in both time and space creates temporal boundaries in the medium and leads to ``time-reflection'' and ``time-refraction''  of light and to the eventual formation  of the photonic time crystal within the modulated 
optical material, offering a new regime of light-matter interactions  and a potential for practical applications, from  non-resonant light amplification to  tunable lasing. However, the conventional approach commonly used for the photonic time crystals and related phenomena  that relies on the concept of effective  time-dependent refractive index, is fundamentally unsuitable to this domain of ultra-fast modulation at optical time scales. We develop the appropriate theoretical description of the electromagnetic response and the resulting phenomena in this regime, and demonstrate not only quantitative but also  qualitative differences from the conclusions obtained using the conventional approach.
\end{abstract}

\maketitle

\section{Introduction}

Recent developments in ultra-high frequency modulation of material properties \cite{modulation1,modulation2,modulation3} opened 
the way to study the interaction of light and matter at the temporal boundaries.\cite{VSMoti}  In an analogy to a spatial interface, a temporal boundary leads to the refraction and reflection of the light propagating in the material, the phenomenon generally referred to as the  ``time-refraction" and ``time-reflection".\cite{time_reflection}  When the material modulation is periodic in time, it  leads to multiple time-reflections and time-refractions, which interfere to form the effective bandstructure for the propagating electromagnetic waves -- the  photonic time-crystal.\cite{Relly2007,Cervantes2009,PTC1,Shalaev2016,Segev2018,VSMotiR1,VSMotiR2}

However, all these theoretical predictions  rely on the use of the effective time-dependent refractive index $n\left(t\right)$ or the corresponding permittivity $\epsilon\left(t\right)$ to account for the effect of the material modulation on the electromagnetic waves in the medium. While certainly appropriate for the low-frequency range when the 
external modulation is slow at the scale of the characteristic time of the electronic response in the material, this assumption becomes increasingly problematic in the opposite limit that is required for the formation of optical time crystals -- when the modulation time scale is comparable to a single optical cycle \cite{Relly2007,Cervantes2009,PTC1,Shalaev2016,Segev2018,VSMotiR1,VSMotiR2} and thus smaller than the electronic relaxation times by more than an order of magnitude.\cite{Ziman}

In the present work we develop the  theoretical description of the electromagnetic response of a medium modulated at the rate that is comparable to the optical frequency scale, based on the first principles. We demonstrate that the resulting predictions for light at or close to the time boundaries in this regime show not only  quantitative but also  qualitative differences from the conclusions obtained using conventional approach based on the concept of time-dependent refractive index.

\section{The Model}

A large variation in the refractive index of an optical material that is necessary to induce noticeable time-reflections,\cite{VSMoti,time_reflection,Relly2007,Cervantes2009,PTC1,Shalaev2016,Segev2018,VSMotiR1,VSMotiR2} generally relies
on introducing a substantial change in the energies of its electrons, whether bound or free.\cite{BornWolf,Boyd,Khurgin-index}  While this can be achieved in many different ways, from mechanical strain \cite{strain} and acousto-optics \cite{acousto-optics} to  thermal effects \cite{thermal} to  carrier density modulation, \cite{electrooptics} the requirement of ultra-fast time scales necessary for the formation of time-boundaries for propagating light, can only be addressed in the approach that relies on a rapid  change in energy distribution of its electrons by a higher-frequency optical pumping. \cite{VSMoti,Israel,EN2023} 

In all such cases, from intra-band generation of hot carriers \cite{Israel} to inducing transient populations in different electronic bands,\cite{EN2023} the resulting effect on light-matter interactions can be described within the framework of time-dependent population densities  in different electronic (sub)bands or (if electronic coherence is essential)  the corresponding density matrices.\cite{LLQM} These can be calculated (and ultimately measured) in the system that is not subject to the optical probe/signal field, and for the purposes of describing optical signal propagation they can be considered as known {\it a priori}. 

\begin{figure}[htbp] 
 \centering
\includegraphics[width=3.25in]{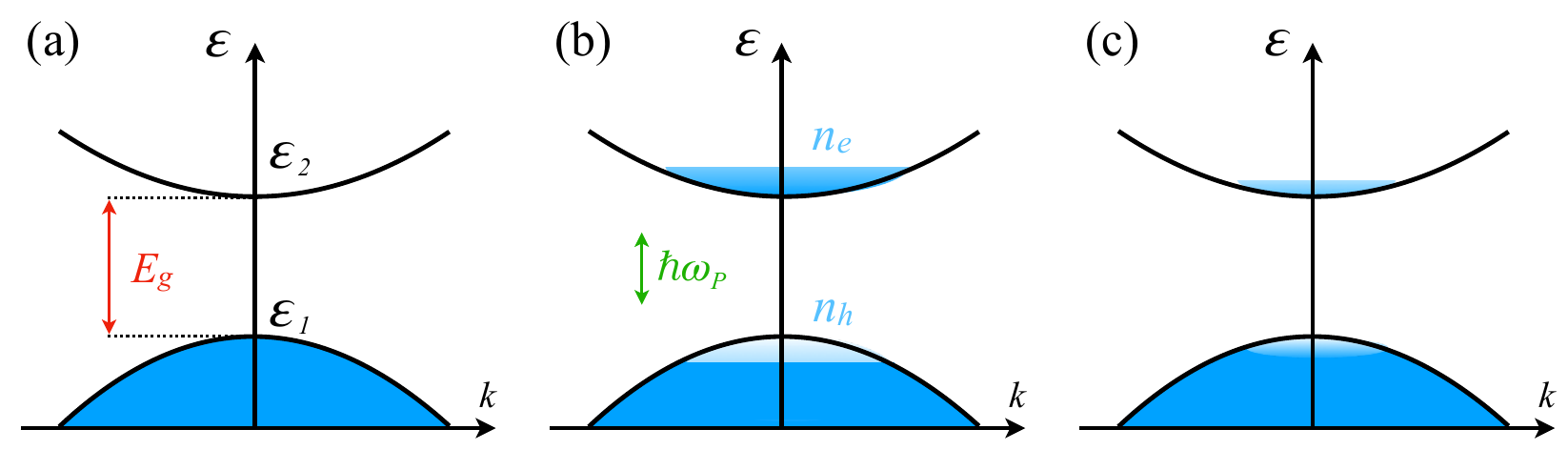} 
   \caption{  The schematic representation of the bandstructure  of a direct bandgap dielectric (or an intrinsic semiconductor), before (a), during (b) and immediately after the optical pump pulse, that induced transient electron and  hole populations. The blue 
   color filling represents the total carriers populations, $\varepsilon_1$ and $\varepsilon_2$ are the band edge energies, 
$ \omega_c \equiv E_g/\hbar$ and $\omega_P$ are respectively the interband absorption edge and the optical pump frequency.   }
   \label{fig:bands}
\end{figure}

In particular, in the case of a dielectric or a wide bandgap  intrinsic semiconductor modulated by virtual transitions \cite{EN2023} due to a strong  optical pump at the frequency below the inter-band absorption cutoff (see Fig. \ref{fig:bands})
\begin{eqnarray}
\hbar \omega_P < E_g \equiv \hbar \omega_c,
\label{eqn:Eg}
\end{eqnarray}
for the induced population densities in the conduction and valence bands we find \cite{EN2023}
\begin{eqnarray}
& & n_e\left(t\right)    = n_h\left(t\right)  =  \left( \frac{e \left| x_{eh} \right| }{2 \pi \hbar}\right)^2 \left(\frac{2}{\hbar} \frac{m_e m_h }{m_e + m_h}\right)^\frac{3}{2} \nonumber \\
& \times & 
\frac{\pi}{\sqrt{\omega_c - \omega_P}}
\left[ I_P\left(t\right) + \frac{e^{- t/\tau_{\rm eh}}}{\tau_{eh}} \int_{-\infty}^t dt' \ I_P\left(t'\right)   \right] ,
\label{eqn:virtual_modulation}
\end{eqnarray}
where $I_P\left(t\right)$ is the intensity of the optical pump with the frequency $\omega_P < \omega_c$, $ex_{eh}$ is the dipole moment matrix element for the (``vertical'' \cite{semiconductor_book}) transition between the centers of the conduction and valence bands, $m_e$ and $m_h$ are the electron and hole  band edge effective masses, and $\tau_{\rm eh}$ is the corresponding  recombination time.

We will therefore formulate the problem addressed in the current work, as finding the electromagnetic response of a material system  under high-frequency modulation, for  { {\it given}  transient population densities induced by the ultra-high frequency modulation.

\section{The Theory}

For free carriers in an optical material subject to an external modulation and a  probe/signal electric field, von Neumann density matrix equation \cite{LLQM} can be expressed as
\begin{eqnarray}
- i \hbar \frac{\partial \hat{\rho}}{\partial t} & = & \left[\hat{ \rho }, {\cal \hat{H}}\right],
\end{eqnarray}
where the effective Hamiltonian
\begin{eqnarray}
{\cal \hat{H}} & = & \hat{H}_e + \hat{H}_\gamma + \hat{H}_M\left(t\right)  +\hat{H}_E\left(t\right),
\end{eqnarray}
includes the contributions of the electron energy in the (periodic) crystal lattice $\hat{H}_E$, the interactions leading to the decoherence of the electronic system $\hat{H}_\gamma$, the (ultra-fast) modulation (whether optical or otherwise) $\hat{H}_M\left(t\right)$, and the electromagnetic interaction with the optical signal (``probe'') field $\hat{H}_E\left(t\right)$. We then obtain  
\begin{eqnarray}
i \hbar \frac{\partial \rho_{n_1 {\bf k}_1, n_2 {\bf k_2}}}{\partial t} & = &\left(  \varepsilon_{n_1{\bf k}_1}  -  \varepsilon_{n_2{\bf k}_2} \right) \rho_{n_1 {\bf k}_1, n_2 {\bf k_2}}\nonumber \\
& - & i \hbar \ \gamma_{n_1 {\bf k}_1, n_2 {\bf k_2}} \left(\rho_{n_1 {\bf k}_1, n_2 {\bf k_2}}- \rho^{(eq)}_{n_1 {\bf k}_1, n_2 {\bf k_2}} \right) \nonumber \\
& - & \left[ \hat{H}_M+  \hat{H}_E, \hat{\rho} \right]_{n_1 {\bf k}_1, n_2 {\bf k_2}} , \ \ \ \ \ \label{eqn:rho}
\end{eqnarray}
where $\hat{\rho}$ is the electron density matrix operator with the equilibrium value $\hat{\rho}^{(eq)}$, $n$ and ${\bf k}$ are the  electron band index and the Bloch momentum (with $n=0$ chosen for the highest occupied band in the low-temperature limit), and $\gamma$ is the matrix of  the corresponding (phenomenological) relaxation coefficients.\cite{Boyd}

While we're interested in the linear response regime for the optical ``probe'' field, the system response to the (ultra-fast) modulation can be essentially non-perturbative. We should therefore solve Eqn. (\ref{eqn:rho}) in the ``mixed'' regime, when the response to the optical probe/signal field
${\bf E}\left({\bf r}, t\right)$ can be treated in a perturbative framework, while the effect of the modulation is incorporated at the exact level. 

Let
$\rho^M$ be the exact solution of Eqn. (\ref{eqn:rho}) in the limit  $\hat{H}_E \to 0$, then in the leading order in the  signal/probe
electric field ${\bf E}\left({\bf r}, t\right) $ 
we obtain
\begin{eqnarray}
\hat{\rho}\left(t\right) & = & \hat{\rho}_M\left(t\right)  + \delta\hat{\rho}\left(t\right) ,
\end{eqnarray}
where
\begin{eqnarray}
\frac{\partial \ \delta \rho_{n_1 {\bf k}_1, n_2 {\bf k_2}}}{\partial t} & = &
- \frac{i}{\hbar}  \left(  \varepsilon_{n_1{\bf k}_1}  -  \varepsilon_{n_2{\bf k}_2} \right) \delta \rho_{n_1 {\bf k}_1, n_2 {\bf k_2}}  \nonumber \\
& - & \ \gamma_{n_1 {\bf k}_1, n_2 {\bf k_2}} \delta \rho_{n_1 {\bf k}_1, n_2 {\bf k_2}}\nonumber \\
& + & 
\frac{i}{\hbar} \left[ \hat{H}_M, \delta \hat{\rho} \right]_{n_1 {\bf k}_1, n_2 {\bf k_2}}  \nonumber \\
& - & e {\bf E}\left({\bf r}, t\right) \delta_{n_1n_2} \delta_{{\bf k}_1 {\bf k}_2} \frac{\partial}{\partial{\bf k_1}} \rho^M_{n_1 {\bf k}_1, n_2 {\bf k_2}}, \label{eqn:drho}
\end{eqnarray}
due to the orthogonality of the Bloch functions from different bands  at the same wavevector ${\bf k}$. Note that in the calculation of the matrix elements of $\hat{H}_E$ in (\ref{eqn:rho}),  one must neglect the spatial variation of
the electromagnetic field, since even for free charge carriers their mean-free path that sets the upper limit on the nonlocality of the resulting electromagnetic response, does not exceed a small fraction of the operating optical wavelength.

As immediately follows from Eqn. (\ref{eqn:drho}), the linear response density matrix 
$\delta\hat{\rho}$ is diagonal, with
\begin{eqnarray}
\frac{\partial \ \delta \rho_{n {\bf k}}}{\partial t} & = & 
 - e {\bf E}\left({\bf r}, t\right)  \frac{\partial  \rho^M_{n{\bf k}}}{\partial {\bf k}}
 - \gamma_{n {\bf k}} \delta \rho_{n {\bf k}} , \label{eqn:drhod}
\end{eqnarray}

The essential feature of strong modulation of the electromagnetic response is a qualitative change in the electron distribution -- from intra-band generation of hot carriers to introducing transient carrier populations due to inter-band transitions. In all such cases, the difference between the density matrix $\hat{\rho}^M$ and its equilibrium value $\hat{\rho}^{(eq)}$ can be represented as a sum over transient contributions in different energy 
sub-bands:
\begin{eqnarray}
\hat{\rho}^M\left(t\right)  & = & \hat{\rho}^{(eq)} + \sum_m \hat{\rho}^{(m)}\left(t\right) ,
\label{eqn:rhom}
\end{eqnarray}
where the index $m=0$ is reserved for the highest (partially) occupied band at equilibrium conditions. Note that the contributions for $m\neq 0$ do not necessarily correspond to individual (other) bands, but instead represent physically distinct carrier groups (e.g. hot electrons in the conduction band, or different  valleys (pockets) of individual bands). From the charge conservation
\begin{eqnarray}
\sum_m {\rm Tr}\left[ {\hat{\rho}}^{(m)}\left(t\right) \right] = 0.
\end{eqnarray}

Introducing the effective average scattering rates for each charge carrier group,
\begin{eqnarray}
\frac{1}{\tau_0} & \equiv & \frac{\sum_{\bf k} \rho^{(eq)}_{0{\bf k}} \gamma_{0,{\bf k}}}{\sum_{\bf k} \rho^{(eq)}_{0{\bf k}} }
\end{eqnarray}
and
\begin{eqnarray}
\gamma_m & \equiv & \frac{\sum_{ {\bf k}} \rho^{(m)}_{{\bf k}} \gamma_{n_m,{\bf k}}}{\sum_{{\bf k}} \rho^{(m)}_{{\bf k}} },
\end{eqnarray}
where $n_m$ is the electronic band supporting the $m$-th transient group, we obtain
\begin{eqnarray}
\delta \hat{\rho}\left(t\right) & = & \hat{\bar{\rho}}\left(t\right) + \sum_m \delta\hat{\rho}^{(m)}\left(t\right) ,
\label{eqn:drhom}
\end{eqnarray}
where
\begin{eqnarray}
 \delta \bar{\rho} & =& - e \hbar {\bf v}_{0,{\bf k}} \frac{\partial \rho^{(eq)}_{0,{\bf k}}}{\partial \varepsilon_{0,{\bf k}}} 
 \int_{-\infty}^{t} dt' \  {\bf E}\left({\bf r},t'\right) e^{ - \frac{t - t'}{\tau_0}},
\end{eqnarray} 
and
\begin{eqnarray}
 \delta \rho^{(m)} & =& - e 
 \int_{-\infty}^{t} dt' \ 
 \frac{\partial\rho^{(m)}_{{\bf k}}}{\partial {\bf k}}
 {\bf  E}\left({\bf r}, t'\right) 
 e^{ - \gamma_m \left(t - t'\right)}.
\end{eqnarray} 
Here $\varepsilon_{n,{\bf k}}$ is the electron energy in the $n$-th band, ${\bf v}_{n,{\bf k}} \equiv \partial \varepsilon_{n,{\bf k}}/\partial\hbar{\bf k}$ is the corresponding group velocity, and $\bar{\rho}$ is the linear response density matrix in the absence of modulation.

Then the current density
\begin{eqnarray}
{\bf j}\left(t\right)& = & {\rm Tr}\left[  \hat{\rho} \ \hat{\bf j}\right] = \bar{\bf j}\left(t\right) + \delta {\bf j}\left(t\right), 
\label{eqn:j}
\end{eqnarray}
where
\begin{eqnarray}
\bar{\bf j} & = & \frac{n_0 e^2}{m_*} \int_{- \infty}^t dt' \ {\bf E}\left({\bf r},t'\right) \ \exp\left({- \frac{t - t'}{\tau_0}}\right),
\label{eqn:j0}
\end{eqnarray}
and
\begin{eqnarray}
 \delta {\bf j} & = &  e^2 \int_{- \infty}^t dt' \  {\bf E}\left({\bf r}, t'\right) \sum_m \frac{n^{(m)} }{m^{(m)}_*}  e^{- \gamma_m \left(t - t'\right)}.
 \label{eqn:dj}
\end{eqnarray}
Here $n_0$ is the equilibrium electron density in the top occupied band
\begin{eqnarray}
n_0 \equiv 2 \sum_{\bf k} \rho_{\bf k}^{(eq)},
\end{eqnarray}
$n^{(m)}\left(t\right)$ is the partial  transient density of the $m$-th group, induced by the material modulation,
\begin{eqnarray}
n^{(m)}\left(t\right) \equiv 2 \sum_{\bf k}  \rho^{(m)}_{{\bf k}}\left(t\right),
\end{eqnarray}
the factor of $2$ comes form the electron spin, 
 the effective masses $m_*$ and  $m_*^{(m)}$ are defined by the transient population averages
\begin{eqnarray}
\frac{1}{m_*} & \equiv & \frac{1}{\hbar}  \ \langle \ \frac{\partial {\bf v}_{0 \bf k}}{\partial {\bf k}}\ \rangle^{(eq)}  \nonumber \\
& = & \frac{1   }{\hbar \sum_{\bf k} \rho^{(eq)}_{0{\bf k}} } \sum_{\bf k}  \frac{\partial {\bf v}_{0 \bf k}}{\partial {\bf k}}\rho^{(eq)}_{0{\bf k}}  \ 
\label{eqn:m0}
\end{eqnarray}
and
\begin{eqnarray}
\frac{1}{m_*^{(m)}} & \equiv & \frac{1}{\hbar}  \ \langle \ \frac{\partial {\bf v}_{n_m \bf k}}{\partial {\bf k}}\ \rangle^{(m)} \nonumber \\
& = & \frac{1}{\hbar \sum_{\bf k} \rho^{(m)}_{{\bf k}}  } \sum_{\bf k}  \frac{\partial {\bf v}_{n_m \bf k}}{\partial {\bf k}}\rho^{(m)}_{{\bf k}}, 
\label{eqn:m}
\end{eqnarray}
and are not necessarily related to the band edge effective mass values. 
Note that when the highest occupied band is completely full, $1/m_* = 0$.

In the general case, the partial effective masses $m_*^{(m)}$ depend on the energy distribution of the induced transient populations, such as e.g.
in the case of the modulation via excitation of hot carriers whose dynamic response is affected the band non-parabolicity. 
But even for a complex modulation format leading to multiple transient populations in different effective bands, 
the energy-resolved transient densities that are fully controlled by the actual time-dependent modulation,  
define all the necessary parameters for the current density of Eqns. (\ref{eqn:j}),(\ref{eqn:j0}),(\ref{eqn:dj}) via the straightforward averaging in (\ref{eqn:m0}),(\ref{eqn:m}).
  
Furthermore, in  many practical cases such as e.g. ultrafast modulation by virtual transitions (see Fig. \ref{fig:bands}  and Eqn. 
(\ref{eqn:virtual_modulation})) or when the transient carriers are excited close to the band extrema in different valleys/pockets, the values of $m_*^{(m)}$ correspond to the band edge effective masses. In particular, for the case of a wide band intrinsic semiconductor  modulated via virtual transitions (Fig. \ref{fig:bands}) we find  $1/m_* = 0$ and  $m_*^{(1)} = -m_h$, $m_*^{(2)} =  m_e$ (see Fig. \ref{fig:bands}), where $m_e$ and $m_h$ are the electron and hole effective masses.

Substituting the current density from Eqns. (\ref{eqn:j}), (\ref{eqn:j0}),(\ref{eqn:dj})  into Maxwell's Equations, for the electric field we obtain
\begin{eqnarray}
\epsilon_\infty  \frac{\partial^2 {\bf E}}{\partial t^2} &  + &  c^2\ {\rm curl} \  {\rm curl} \ {\bf E}   + 
 \frac{4 \pi e^2 n_0}{m_*}  {\bf E} 
\nonumber \\
& + &  \left[ {4 \pi e^2} \sum_m \frac{n^{(m)}\left({\bf r},t\right)}{m^{(m)}_*}  \right]  {\bf E} \nonumber \\
& = & \int_{-\infty}^t dt' \ K\left({\bf r}, t, t'\right) \ {\bf E}\left({\bf r}, t' \right),
\label{eqn:wave1}
\end{eqnarray}
where
\begin{eqnarray}
K\left({\bf r}, t, t'\right)  & = &  4 \pi e^2 \left[ \frac{ n_0}{m_* \tau_0} \ \exp\left({- \frac{t - t'}{\tau_0}}\right) \right.  \nonumber \\
& + & \left.  \sum_m \frac{\gamma_m }{m^{(m)}_*} \ n^{(m)}\left({\bf r},t\right)
e^{- \gamma_m \left(t - t'\right)} \right],
\label{eqn:K}
\end{eqnarray}
and $\epsilon_\infty$ is the ``background'' contribution to permittivity of the material (due to e.g. inner core electrons), that is not affected by the modulation.

The wave equation (\ref{eqn:wave1})  offers a complete description of the evolution of the electromagnetic field ${\bf E}\left({\bf r}, t\right)$ for a  given spatiotemporal variation of the transient carrier populations $n^{(m)}\left({\bf r}, t\right)$, and represents the main result of the present work.

When the total duration of the modulation cycle $T$ is well below all intra-band carrier relaxation times,
\begin{eqnarray}
T \ll \tau_0, {1}/{\gamma_m},
\label{eqn:T} 
\end{eqnarray} 
the nonlocal term in the right-hands side of Eqn. (\ref{eqn:wave1}) which scales as $\sim T/\tau_0$ and as $ \sim \gamma T$, can be neglected, reducing our wave equation to
\begin{eqnarray}
  \frac{\partial^2 {\bf E}}{\partial t^2}      
  & + & \frac{4 \pi e^2 n_0}{\epsilon_\infty }   \left[ \frac{ n_0}{m_*}  +  \sum_m \frac{n^{(m)}\left({\bf r},t\right)}{m^{(m)}_*}  \right]  {\bf E} 
 \nonumber \\
 & = & -   \frac{c^2}{\epsilon_\infty} \ {\rm curl} \  {\rm curl} \ {\bf E}.
\end{eqnarray}

On the other hand, for the important special case of a plane wave
\begin{eqnarray}
{\bf E}\left({\bf r}, t\right) & = &E_0   {\bf \hat{e}} \exp\left(i q z - \i \omega_0 t \right), \ \ \ {\bf \hat{e}}\perp\hat{\bf z},
\label{eqn:Ex}
\end{eqnarray} 
 initially propagating in an isotropic material that is subject to spatially uniform temporal modulation, via the Emmy Noether's theorem 
 \cite{Noether} the translational invariance of the system preserves the polarization and the  spatial dependence of the electric field for all times, and
the wave equation  (\ref{eqn:wave1})  reduces to 
\begin{eqnarray}
  \frac{d^2 { E}}{d t^2} 
 & + &  \left[ \omega_0^2 + \frac{4 \pi e^2}{\epsilon_\infty} 
  \sum_m \frac{n^{(m)}\left(t\right)}{m^{(m)}_*} 
  \right]  { E} \nonumber \\
& = &\frac{1}{\epsilon_\infty} \int_{-\infty}^t dt'  K\left(t, t'\right) E\left( t' \right),
\label{eqn:wave4}
\end{eqnarray}
or equivalently
\begin{eqnarray}
  \frac{d^2 { E}}{d t^2} 
  +   \left[ \omega_0^2 +\Omega\left(t\right)^2
  \right]  { E} \
= \int_{-\infty}^t \frac{dt'}{\epsilon_\infty}   K\left(t, t'\right) E\left( t' \right), \ \ \ \ 
\label{eqn:wave5}
\end{eqnarray}
where
\begin{eqnarray}
\Omega\left(t\right) \equiv \sqrt{\frac{4 \pi e^2}{\epsilon_\infty} 
  \sum_m \frac{n^{(m)}\left(t\right)}{m^{(m)}_*} }.
\end{eqnarray}

\section{Discussion}

With the presence of the temporal nonlocality due to the time integral in  ({\ref{eqn:wave1}) and its limiting case (\ref{eqn:wave5}) for plane wave propagation, our  approach is substantially different from the conventional framework based on time-modulated refractive index $n\left({\bf r},t\right)$ or equivalently the time-dependent dielectric permittivity $\epsilon\left({\bf r},t\right) \equiv n\left({\bf r},t\right)^2$, leading to the wave equation
\begin{eqnarray}
{\rm curl} \ {\rm curl} \ {\bf E}\left({\bf r}, t\right) 
 + \frac{1}{c^2} \frac{\partial^2}{\partial t^2} \epsilon\left({\bf r}, t\right) {\bf E}\left({\bf r}, t\right) & = & 0,
 \label{eqn:wave_n}
\end{eqnarray}
that for plane wave propagation in a uniform isotropic  medium reduces to  
\begin{eqnarray}
\frac{d}{dt^2} \epsilon\left(t\right) E\left(t\right) + \epsilon_\infty \omega_0^2 E\left(t\right)  & = & 0.
\label{eqn:wave_n1}
\end{eqnarray}
This conventional approach therefore completely neglects the essential temporal nonlocality of the actual 
carrier dynamics subject to ultra-fast modulation, that is accurately described by Eqn. (\ref{eqn:wave1},(\ref{eqn:wave5}).

The essential difference between the conventional approach based on the time dependent refractive index (permittivity) leading to 
(\ref{eqn:wave_n}) and (\ref{eqn:wave_n1}) and our result (\ref{eqn:wave1}),(\ref{eqn:wave5}), however does not reduce to the treatment of the inherent temporal nonlocality in the material, but has more profound
origins and consequences. To uncover this behavior, we will consider the limiting case when the modulation time scale $T$ is much shorter than the relaxation time of the electronic system (\ref{eqn:T}) and the temporal nonlocality in (\ref{eqn:wave1}) can be neglected. The resulting wave equation then
shows the appearance of artificial similarity to (\ref{eqn:wave_n}) and (\ref{eqn:wave_n1}). In particular, for  plane wave propagation in a uniform isotropic medium (\ref{eqn:Ex}) in both cases we find the effective 
Schr\"odinger's equation
\begin{eqnarray}
- \frac{d\psi\left(t\right)}{dt^2} + V_{\rm eff}\left(t\right)   \psi\left(t\right) & = & \varepsilon \psi\left(t\right),
\end{eqnarray}
where the effective ``energy''$\varepsilon$  is defined by the square of the electromagnetic wave frequency before the modulation,
\begin{eqnarray}
\varepsilon \equiv \omega_0^2,
\end{eqnarray}
and the ``wavefunction'' $\psi$  is  defined as the time-dependent electric field  for the actual dynamics of (\ref{eqn:wave_n}),(\ref{eqn:wave_n1}), 
\begin{eqnarray}
\psi\left(t\right) \to E\left(t\right),
\label{eqn:psi}
\end{eqnarray}
and as the electrical displacement  for the conventional time-dependent refractive index approach,
\begin{eqnarray}
\psi\left(t\right) \to D\left(t\right) \equiv \epsilon\left(t\right) E\left( t \right),
\label{eqn:psin}
\end{eqnarray}
In both cases, the effective potential $V_{\rm eff}\left( t\right)$ is defined by the material modulation; for the exact dynamics
\begin{eqnarray}
V{\rm eff} & = & - \Omega\left(t\right)^2 \equiv \frac{4 \pi e^2}{\epsilon_\infty} 
  \sum_m \frac{n^{(m)}\left(t\right)}{m^{(m)}_*},
  \label{eqn:V}
\end{eqnarray}
while for the conventional approach
\begin{eqnarray}
V_{\rm eff} & = & - \omega_0^2\left( \frac{\epsilon_\infty}{\epsilon\left(t\right)} - 1\right). 
\label{eqn:Vn}
\end{eqnarray}

Despite the major difference in the  physical meaning of the wavefunction $\psi\left(t\right)$ (see Eqns. (\ref{eqn:psi}), (\ref{eqn:psin})), 
with the reduction to effective Schr\"odinger's equation in both cases it may be tempting to assume, at least on a qualitative level,  that
the use of the time-dependent refraction index is appropriate, as 
one can use the ``effective'' permittivity that leads to the same time dependence in the effective potential $V_{\rm eff}$ as the actual  dynamics of Eqn. (\ref{eqn:V}). However, there is glaring and qualitative difference in the effective potentials of (\ref{eqn:V}) and (\ref{eqn:Vn}): while the latter is
proportional to the square of the actual (optical) frequency  of the propagating wave, the former does not explicitly depend on its frequency at all.

To illustrate the resulting qualitative difference between the predictions of the time-dependent index approach and the actual dynamics, we consider the case of optical modulation by a soliton ``pump'' pulse, leading to the $\sim 1/\cosh^2\left(t/T\right)$ time dependence of the effective modulation potential $V_{\rm eff}\left(t\right)$, with
\begin{eqnarray}
V_{\rm eff} & = & \frac{\Omega_0^2}{\cosh^2\left(t/T\right)}.
\label{eqn:PT}
\end{eqnarray} 
In the case of the time-dependent index approach $\Omega_0$ is proportional to the initial frequency of the propagating wave, while
for the exact dynamics it's frequency independent and scales with the maximum density of the transient 
carrier population induced by the pump.

This results in the scattering problem for the modified P\"oschl-Teller potential well, which has the exact analytical solution \cite{LLQM}
\begin{eqnarray}
\lim_{\left|t\right| \gg T}  \psi\left(t\right) & = & \left\{
\begin{array}{cc}
e^{i q z - i \omega_0 t }, & t < 0,  \\
t \ e^{i q z - i \omega_0 t } + r \ e^{i q z +  i \omega_0 t } , & t >0,
\end{array}
\right.
\end{eqnarray}
where the wavenumber
\begin{eqnarray}
q & = &  \frac{\omega_0}{c} \sqrt{\epsilon_\infty}, 
\end{eqnarray}
 the time-reflection coefficient
\begin{eqnarray}
\left| r \right|^2 & = & \frac{\cos^2\left[\frac{\pi}{2} \sqrt{1 + \Omega_0^2 T^2 }\right]}{\sinh^2\left(\pi \omega_0 T \right)},
\end{eqnarray}
and the time-transmission coefficient
\begin{eqnarray}
\left| r \right|^2 & = & 1 + \left| r \right|^2.
\end{eqnarray}
Note that for 
\begin{eqnarray}
\Omega_0 & = & \frac{2}{T} \sqrt{ \ell \left( \ell + 1\right) },
\label{eqn:refless}
\end{eqnarray}
where $\ell$ is a positive integer, the time-reflection coefficient is exactly zero -- a property that is well known for the scattering by the
modified P\"oschl-Teller potential well. \cite{explanation} This however has very  different implications for the exact system 
dynamics and for the time-dependent index description.

\begin{figure}[htbp] 
 \centering
\includegraphics[width=3.45in]{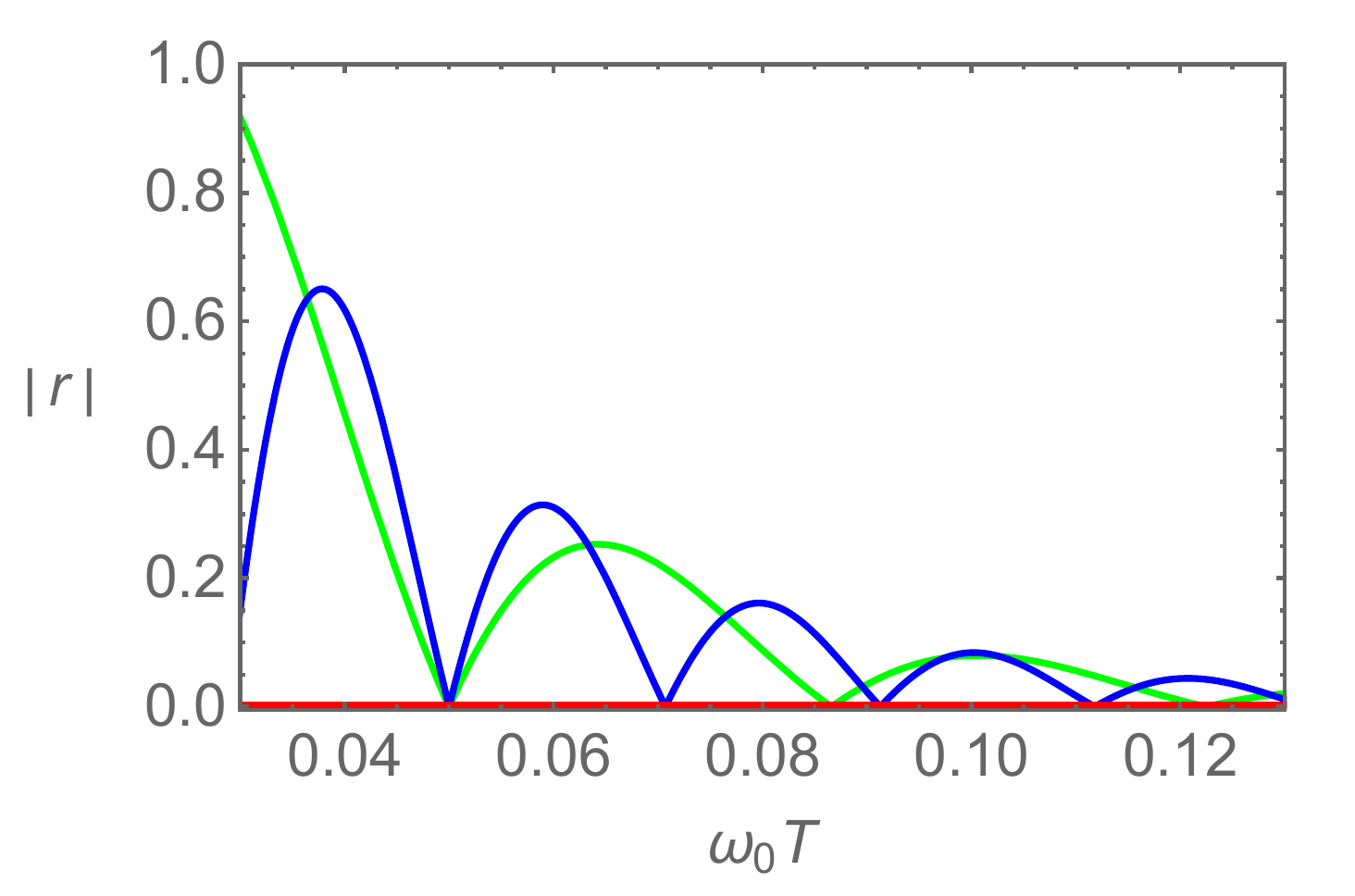} 
   \caption{  The magnitude of the time-reflection coefficient $|t\left(\omega_0\right)$ as a function of the initial frequency of the propagating wave $\omega_0$, for scattering by a uniformly time-modulated medium
   with the   modified P\"oschl-Teller time variation, Eqns. (\ref{eqn:PT}), (\ref{eqn:refless}), for $\ell = 1$ (green), and $\ell = 2$ (blue).
  Red line represents the actual  zero time-reflection  for all these parameters, while solid curves show the corresponding results of the time-dependent refractive index approximation.  }
   \label{fig:time-reflection}
\end{figure}
In the exact picture, $\Omega_0$ is frequency-independent, and when the modulation amplitude satisfies the condition (\ref{eqn:refless}), 
the system shows no time-reflection for {\it all} signal frequencies $\omega_0$. In a dramatic contrast to this behavior, due to the 
 frequency dependence of  $\Omega_0 \propto \omega_0$ for the time-dependent refractive index approach (\ref{eqn:Vn}),(\ref{eqn:PT}), Eqn. (\ref{eqn:refless}) can only be satisfied for {\it distinct} values of the signal frequency, leading to noticeable time-reflection otherwise. This key difference in behavior is illustrated in Fig. \ref{fig:time-reflection}.

This qualitative difference of the actual lightwave dynamics in time-modulated media at optical frequencies from the predictions of the approach 
based on the time-dependent index of refraction, puts to question the extension of  its results to the optical domain, and especially to those related
to photonic time crystals, as the latter concept strongly relies on constructive and destructive interference due to multiple time-reflections. While the  mapping to the effective Shr\"odinger's equation may protect at least some of the existing results of the effective time-dependent index approach in the optical domain, this is in no way ensured, and they must be carefully re-examined.

\section{Conclusions}

We have developed the theoretical description of the electromagnetic response and the resulting phenomena in the regime of ultra-high frequency modulation of the material properties, and demonstrated that the resulting behavior shows  not only quantitative but also  qualitative differences from the conclusions obtained using the conventional approach based on time-dependent refractive index model. 



\begin{thebibliography}{10}
\bibitem{explanation} For this reason the modified P\"oschl-Teller profile (\ref{eqn:PT}) is frequently referred to  as the ``reflectionless potential'', \cite{LLQM} and
its functional dependence in the form of a fundamental soliton pulse \cite{soliton}  has a profound physical meaning as  the manifestation of the inverse-scattering theory \cite{inverse_scattering}   that connects the soliton solutions of the nonlinear Schr\"odinger's equation to a special class of 
the linear scattering problem.
\end{thebibliography}
\end{document}